\documentclass{emulateapj}

\begin{document}

\title{A Dark Energy Camera Search for an Optical Counterpart to the
  First Advanced LIGO Gravitational Wave Event GW150914}

\shorttitle{DECam Search for an Optical Counterpart to GW150914}

\shortauthors{Soares-Santos et al.}


\author{
M.~Soares-Santos\altaffilmark{1},
R.~Kessler\altaffilmark{2},
E.~Berger\altaffilmark{3},
J.~Annis\altaffilmark{1},
D.~Brout\altaffilmark{4},
E.~Buckley-Geer\altaffilmark{1},
H.~Chen\altaffilmark{2},
P.~S.~Cowperthwaite\altaffilmark{3},
H.~T.~Diehl\altaffilmark{1},
Z.~Doctor\altaffilmark{2}
% ...
\\ \vspace{0.2cm} (The DES Collaboration) \\
}
 
\altaffiltext{1}{Fermi National Accelerator Laboratory, P. O. Box 500, Batavia, IL 60510, USA}
\altaffiltext{2}{Kavli Institute for Cosmological Physics, University of Chicago, Chicago, IL 60637, USA}
\altaffiltext{3}{Harvard-Smithsonian Center for Astrophysics, 60 Garden Street, Cambridge, MA, 02138}
\altaffiltext{4}{Department of Physics and Astronomy, University of Pennsylvania, Philadelphia, PA 19104, USA}
% ...

\begin{abstract}
  We report initial results of a deep search for an optical counterpart to
  the gravitational wave event GW150914, the first trigger
  from the Advanced LIGO gravitational wave detectors.
  We used the Dark Energy Camera (DECam) to image
  a 102 deg$^2$ area, corresponding to $38\%$ of the initial trigger
  high-probability sky region and to $11\%$ of the revised
  high-probability region. 
  We observed
  in $i$ and $z$ bands at 4--5, 7, and 24 days after the
  trigger.  
  The median $5\sigma$
  point-source limiting magnitudes of our search images are $i=22.5$
  and $z=21.8$ mag. 
  We processed the images through a difference-imaging
  pipeline using templates from 
  pre-existing Dark Energy Survey data and publicly
  available DECam data. 
  Due to missing template observations and other losses, 
  our effective search area subtends 40 deg$^{2}$,
  corresponding to 12\% total 
  probability in the initial map and 3\% of the final map.   
  In this area, 
  we search for objects that decline significantly between days 4--5 
and day 7,
and are undetectable by day 24, finding none to typical magnitude
limits of $i= 21.5,21.1,20.1$ for object colors $(i-z)=1,0,-1$, respectively.
  Our search demonstrates the feasibility of a dedicated
  search program with DECam and bodes well for future research in this
  emerging field.
\end{abstract}


\section{Introduction}

The advanced network of ground-based gravitational wave (GW)
interferometers is designed to detect and study GW emission from
events such as the mergers of binary systems composed of neutron stars
and/or black holes to distances of hundreds of Mpc (see
\citet{ligo2013} and references therein).  In mergers containing at
least one neutron star, counterpart electromagnetic radiation is expected,
potentially ranging from a short-duration gamma-ray burst through
optical/near-IR emission from the radioactive decay of r-process
nuclei to radio emission from ejecta interacting with the circumbinary
medium (e.g.,
\citealt{li98,nakar2011,metzger2012,barnes2013,2013ApJ...775..113T,2014ApJ...780...31T,2014ApJS..211....7A,berger2014,cowperthwaite2015}).
The detection of an electromagnetic counterpart will provide critical
insight into the physics of the event, 
helping to determine the distance scale, energy scale, and the
progenitor environment, as well as insight into the behavior of matter
post-merger (e.g., the production of jets and outflows).

With this motivation, we recently began an observational program
using the wide-field Dark Energy Camera (DECam; \citealt{flaugher2015}) on the
Blanco 4-m telescope at Cerro Tololo Inter-American Observatory to
search for optical counterparts to GW triggers from the new advanced
GW detectors (LIGO, \citet{ligo2009}; Virgo, \cite{virgo2009}).
This program was awarded three target of opportunity nights to observe LIGO-triggered events during the 2015B semester; observations were coordinated with and managed by the Dark Energy Survey (DES).
Our program is optimized for detection of kilonovae, the hypothesized 
optical counterparts
of mergers involving neutron stars, which would appear as
red transients with expected 
decay timescale of about a week  (for an overview of our program see \citet{des2016}).

On 2015 September 14 at 09:50:45 UT the Advanced LIGO interferometer network detected
a high significance candidate GW event designated GW150914 
\citep{ligo2016} and two
days later provided spatial location information in the form of
probability sky maps via a private GCN circular (\#18330; \cite{GCN18330}).
We initiated observations with DECam, a 3 deg$^2$ field-of-view 
instrument, on 2015
September 18 in an effort to identify an optical counterpart.  Here
we describe the observations and provide the results of the
three-epoch search.  These DECam observations  are
the deepest search for an optical counterpart to GW event GW150914 
\citep{gw-em-overivew}.

\section{DECam Observations of GW150914}
\label{sec:data}

The detection of GW150914 was triggered by the cWB (coherent
WaveBurst; \citet{2008CQGra..25k4029K}) 
unmodeled burst analysis during real-time data processing.
On 2015 September 16, the LIGO Virgo Collaboration (LVC) provided two
all-sky localization probability maps for the event, generated from
the cWB and LALInferenceBurst (LIB; \citet{veitch2014}) analyses.  
The cWB online trigger
analysis makes minimal assumptions about signal morphology by
searching for coherent power across the LIGO network.  The LIB
analysis is a version of the LALInference analysis Bayesian
forward-modeling-based follow up tool that uses a Sine-Gaussian signal
morphology instead of models of compact binary mergers
\citep{veitch2014}; for information on both algorithms see
\citet{essick2015}.  The maps provided initial spatial localization
of 50\% and 90\% confidence regions encompassing about 200 and 750
deg$^2$, respectively.

Our first observations with DECam took place on 2015
September 18 UT.  Overall, we imaged 102 deg$^2$ covering
38\% of the total probability in the initial cWB map;
see Table~\ref{datatable} for a summary of our DECam observations.
As shown in Figure~\ref{datafig}, 18 deg$^2$, were centered on the LMC.
For the remaining 84 deg$^2$
we obtained 3 separate epochs of imaging. 
At each epoch we acquired one
   90-sec exposure in $i$ band  and two 90-sec exposures
   in $z$ band. 
  The first
epoch spanned 4--5 days post-GW trigger (2015 September 18--19 UT),
the second epoch 7 days post-GW trigger (2015 September 21 UT), and the
third was obtained 24 days post-GW trigger (2015 October 08 UT).

Subsequently, in January 2016, the LVC released a revised sky map of
localization probabilities from a LALInference analysis that used the
assumption that the signal arises from a compact binary coalescence
(CBC). That analysis also showed that the 
data are most consistent with models of a binary black hole merger
(BBH). The LALInference-based map is considered the most accurate and
authoritative localization for this event.  
Our 102 deg$^2$  cover a total of $11\%$ probability in this new map,
as the localization region has 
shifted significantly southward (see Figure~\ref{datafig}) relative
to the initial cWB map.

Our single-epoch exposures achieve median $5\sigma$ point-source
limiting magnitudes of $i = 22.5$ and $z = 21.8$ with an rms variation
among the images of $\pm 0.5$ mag.  This value is a consequence of
night-to-night variations in the observing conditions (see
Table~\ref{datatable}) and of a strong gradient in stellar density and
extinction along the major axis of the region imaged (see
Figure~\ref{datafig}).

\begin{deluxetable*}{lrrrccccc}
\tabletypesize{\footnotesize}
\tablecaption{Summary of Observations 
\label{datatable}}
\tablehead{
    \colhead{Program} &  
    \colhead{Night}   & 
    \colhead{MJD}     &
    \colhead{$\Delta t$\tablenotemark{a}} &
    \colhead{$\langle$PSF(FWHM)$_i$$\rangle$}    &
    \colhead{$\langle$airmass$\rangle$}  &
    \colhead{$\langle$depth$_i\rangle$}   &
    \colhead{$\langle$depth$_z\rangle$}  &
    \colhead{$A_{\mathrm{eff}}$\tablenotemark{b}} 
\\
    \colhead{ } &
    \colhead{(UT)} & 
    \colhead{ } &
    \colhead{(days)} &
    \colhead{(arcsec)} &
    \colhead{ } &
    \colhead{(mag)} &
    \colhead{(mag)} &
    \colhead{(deg$^2$)} 
}
\startdata
Main, 1$^{\mathrm{st}}$ epoch & 2015-09-18 & 57383 &  3.88 & 1.38 & 1.50 & 22.71 & 22.00 & 52.8 \\      
                     & 2015-09-19 & 57384 &  4.97 & 1.35 & 1.46 & 22.82 & 22.12 & 14.4 \\      
Main, 2$^{\mathrm{nd}}$ epoch & 2015-09-21 & 57286 &  6.86 & 2.17 & 1.51 & 22.18 & 21.48 & 67.2 \\      
Main, 3$^{\mathrm{rd}}$ epoch & 2015-10-08 & 57303 & 23.84 & 1.46 & 1.40 & 22.33 & 21.63 & 67.2 \\[2pt]      
\tableline \\[-5pt]
LMC, initial           & 2015-09-18 & 57383 &  3.98 & 1.14 & 1.30 & 21.32 & 20.62 & 14.4 \\      
LMC, extension    & 2015-09-27 & 57292 & 12.96 & 1.21 & 1.28 & 20.91 & 20.21 & 33.6 \\[-5pt]      
\enddata
\tablecomments{Summary of the observations performed in the ``main''
  search program, described in this paper, and the ``LMC'' program,
  described in the companion paper \citet{Annis:2016}.  We observed at
  high airmass because the region of interest was rising at the
  end of the night. The PSF FWHM, and therefore the actual depth
  achieved, are partly affected by these high airmass conditions.
  The reported depth corresponding to 5-$\sigma$ point source detection in the search images.
  Variations in cloud conditions are also responsible for the
  variation in depth.  The effective area
  imaged in the main program corresponds to 28 camera fields. The area
  covered in the LMC program totaled 20 fields.}
\tablenotetext{a}{Time elapsed between the trigger time and the time
  stamp of the first image of the night.}  \tablenotetext{b}{Effective
  area imaged, considering that approximately 20\% of the 3 deg$^2$
  field of view of DECam is lost due to chip gaps (10\%), 3 dead CCDs
  (5\%) and masked edge pixels (5\%).
}
\end{deluxetable*}

\subsection{Observing Strategy}

We chose the location and sequence of DECam observations using an
automated observing strategy algorithm. 
The algorithm utilizes the GW localization map, an
estimate of the event distance, and a model of the expected optical
emission (e.g., \citealt{barnes2013}).  This information is folded in
with observational information, including a map of sky brightness
(using the DES sky brightness model; 
\citealt{2012ASPC..461..201N}), the atmospheric transmission (using
information on airmass and the interstellar dust extinction from
Planck; \citealt{planck2014}), the expected seeing (from scaling laws
with airmass and wavelength), and the confusion-limit probability
(based on stellar density maps) to produce a full source-detection
probability as a function of sky location.  
We used this map to observe the highest probability region that 
    included area both inside and outside the DES footprint.

\begin{figure}[h!]
  \includegraphics[width=\linewidth,,trim=10 0 10 0, clip]{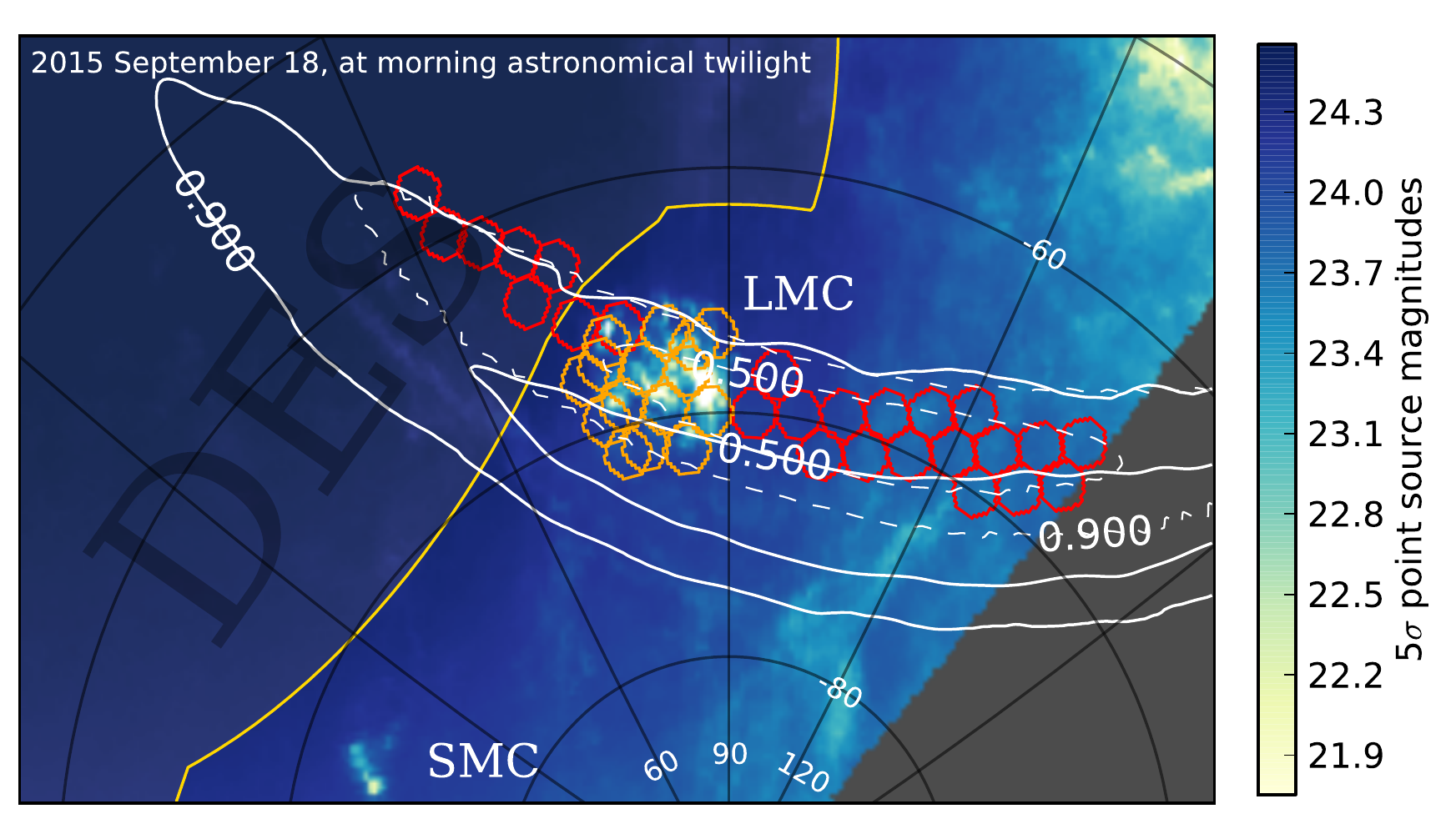}
  \caption{The color image shows the estimated limiting point-source
    magnitude for a 90-sec $i$ band exposure as a function of sky
    position for our first night of DECam observations just before
    sunrise.  In this area and for this time of night, the variations
    are mostly due to interstellar dust extinction.  The dotted
    contours show the initial (September 2015) {\tt
      skyprobcc\_cWB\_complete} map, while the solid contours are for
    the final (January 2016) {\tt LALInference\_skymap}.  There is an
    island of significant probability in the Northern hemisphere in
    the {\tt skyprobcc\_cWB\_complete} map, not present in the {\tt
      LALInference\_skymap}, so the dotted contours do not show the
    complete 50\% or 90\% areas.  The hexagonal DECam fields observed
    are shown, with red for the main search and orange for the short
    exposure LMC data.  Fields located on the west (left) side of the
    region of interest overlap with the DES area (footprint boundary
    shown in light-gold).  The excluded region (dark grey) is beyond
    the horizon limit that could be observed with DECam at that time.
    The total area inside the camera pointings is about 102
    deg$^2$. We covered about 11\% of the total localization
    probability in the final map, and 38\% of the initial map.  
    The projection shown is an equal-area McBryde-Thomas flat-polar quartic projection.}
 \label{datafig}
\end{figure}

In the case of GW150914 the localization region intersected the
Large Magellanic Cloud (LMC), so we designed a separate set of short
observations to observe the brightest LMC stars.  We
obtained 5-sec $i$ and $z$ band exposures covering 18 deg$^2$ centered
on the LMC on 2015 September 18 and 27.  This shallower data set was used to
search for a potental failed supernova in the LMC; the
results are reported in a separate paper \citep{Annis:2016}.
Figure~\ref{datafig} shows a sky map computed for the end of the first
night of observations, zoomed in to the region of interest and
detailing the fields observed in each of the three epochs in red.

\subsection{Image Processing}

Our data analysis relies on subtracting earlier template images from
the science images taken for this program.  In the area that overlaps
the DES footprint (25\% of the total), we used DES images from the
first two seasons of the survey as templates. 
In the 75\% of the area outside of the DES footprint, we used publicly
available DECam data from the NOAO Science Archive
({\tt\url{portal-nvo.noao.edu}}), requiring exposures of at least 30
sec in $i$ and $z$ bands.

We processed the DECam search and template images using the DES Data
Management single-epoch image processing software
\citep{2012ApJ...757...83D,DESDM2012,sevilla2011,Gruendl:2016}. Its output images were
 used as input to the difference imaging pipeline, which we
developed from the DES Supernova pipeline \citep{kessler2015}.  The
main adaptation of the pipeline for our purposes was to generalize to
the case of search and template images with arbitrary relative 
alignment. 
A candidate requires two SExtractor \citep{1996A&AS..117..393B} 
detections in 
the first epoch in both $i$ and $z$ bands. 
To reduce the large number of detected artifacts, 
     each detection must satisfy quality requirements
     (Table 3 of \citealt{kessler2015}) and be selected by our automated scanning 
     program \citep{2015AJ....150...82G}.
     For each of the 2349 candidate locations, ``forced'' PSF-fitted 
     fluxes and uncertainties are obtained at every epoch regardless
     of whether or not there was a detection. 

\section{Analysis}\label{sec:analysis}

While a BBH merger is not expected to result in an optical signature,
it is nevertheless of interest to search for a possible optical counterpart.
As our first epoch of observations occurred 4
days after the trigger, our prior on the search is that any candidate
shall be fading slowly enough to be detectable 7 days after the event,
but not 24 days after the event.  

Of the 84 deg$^2$ area outside of the LMC, about 20\% is lost due to 
camera fill-factor (see Table \ref{datatable} for details) resulting in 
an effective area of 
67.2 deg$^2$. In addition, 
30\% of the area is lost due to 
sparse availability of templates outside of the DES footprint.
Another 10\% loss arises from  processing issues.
This results in 40 deg$^2$ which were used in this analysis.

Based on an analysis of a sample of fake point sources injected into
the images in this area, we find that the typical $80\%$ source detection 
completeness in the subtracted images is at $i\approx 22.1$ and 
$z\approx 21.2$ mag. In the first epoch, where the observing conditions
were better, we achieve that level of completeness at $i\approx 22.7$
and $z\approx 21.8$, comparable to the 5$\sigma$ point source depth for 
those images. 
The fakes were in all the images we processed,
thus the completeness depth reflects the variation in conditions as well.

\subsection{Sample Selection}

For the selection criteria described below, multiple observations 
   per night (primarily in $z$ band) are combined into a single 
   weighted-average flux:

\begin{enumerate}

\item Second-epoch signal-to-noise ratio (S/N) above 2 in both $i$ and $z$ 
(to enable flux change determination with respect to the first epoch);

\item $\ge 3\sigma$ decline in both $i$ and $z$ fluxes from the first epoch to the second (to
  isolate fading sources; $\sigma$ is defined by the quadrature sum of the 
  flux errors in the first two epochs);

\item  S/N $\le 3\sigma$  in both $i$ and $z$ third epoch 
(at 24 days post-trigger, to reject long-timescale transients such as supernovae).

\end{enumerate}

\subsection{Results}

In Table~\ref{cuts} we show the impact of our selection criteria on
the sample of candidates as a function of the first epoch $i$-band magnitude.
The decaying light curve requirement has the most impact in reducing the sample
size. None of the candidates pass all the selection criteria. 
The area analyzed, 40 deg$^2$, covers
3\% of the localization probability in the final LALInference map
(though it covered 12\% in the initial cWB map). 

To interpret these results some caveats are required.
Because our selection criteria impose demands on significance
in the second epoch, the actual first epoch search depth  depends
on the decline rate and $i-z$ color of the source model.  
In addition, we have not yet accounted for the degraded sensitivity 
to candidates located in bright galaxies. 

For a particular source model, we can estimate the search
depth. We applied our selection criteria to a sample of 
fake sources randomly placed in our search images before processing
with our difference imaging pipeline. 
The fakes have a constant decay rate of 0.3 mags/day 
and are red, with $(i-z) \approx 1$, as expected from kilonova models. 
The magnitude at which we recover 
50\% of the fakes, $m_{50\%}$, is about 1 magnitude brighter 
than the 5$\sigma$ point source limiting
magnitude reported in Table~\ref{datatable}, 
i.e., $m_{50\%} - m_{5\sigma} \approx -1$.
Simulations with bluer models show that
for sources  with $(i-z) = 0$ the search depth
is $m_{50\%} - m_{5\sigma}  \approx -1.4$; for  
 $(i-z) = -1$, the search depth is 
$m_{50\%} - m_{5\sigma}  \approx -2.4$.
We therefore achieve magnitude limit of $i= 21.5,21.1,20.1$ for object
colors $(i-z)=1,0,-1$, respectively.

\begin{deluxetable}{lrrrr}
\tablecolumns{5}
\tablecaption{Number of selected events \label{cuts}}
\tablehead{
    \colhead{mag($i$)} &
    \colhead{raw} & \colhead{cut 1} &
    \colhead{cut 2} & \colhead{cut 3} }
\startdata
18.0--18.5 &   84  &   1  & 0 & 0 \\
18.5--19.0 &   177  &   1  & 0 & 0 \\
19.0--19.5 &   291 &   2  & 0 & 0 \\
19.5--20.0 &   227  &   2  & 1 & 0 \\
20.0--20.5 &   156 &   17  & 2 & 0 \\
20.5--21.0 &   225 &  42  & 3 & 0 \\
21.0--21.5 &   334  & 84  & 2 & 0 \\
21.5--22.0 &   756 & 159  & 1 & 0 \\
22.0--22.5 &  1099 & 183  & 0 & 0 \\
\tableline\\[-5pt]
total      &  2349 & 491 & 9 & 0 \\[-5pt]
%
%
\enddata
\end{deluxetable}

\section{Conclusions}
\label{sec:conclusion}

We presented our search for an optical counterpart to the first
gravitational wave event, GW150914, using the wide-field
DECam instrument.  Our observations cover $102$ deg$^2$ corresponding to 11\%
of the total probability map. 
The search images used in this analysis reach median 5$\sigma$ 
point source depth of $i=22.5$ and $z=21.8$
mag.  
Our DECam/Blanco observations are the deepest optical follow-up 
for this GW event.

Using  selection criteria which
isolate fading transients over the analysis region covering
3\% of the total localization probability, 
we find no candidate counterparts.
We are still investigating improved background
rejection criteria using information such as: 
matching against a galaxy catalog 
to remove transients associated with high-redshift galaxies,
angular separation between $i$ and $z$ exposures to reduce 
asteroids, and detailed simulations of supernovae 
and source models 
    to better optimize selection requirements as well
    as the search strategy for future events.

Although these results are not
surprising given the partial areal coverage and the likely
BBH merger nature of the
event, our search is a crucial first step and 
demonstrates the viability of DECam
for deep optical follow-up of GW events.  


\acknowledgments

Funding for the DES Projects has been provided by the U.S. Department of Energy, the U.S. National Science Foundation, the Ministry of Science and Education of Spain, 
the Science and Technology Facilities Council of the United Kingdom, the Higher Education Funding Council for England, the National Center for Supercomputing 
Applications at the University of Illinois at Urbana-Champaign, the Kavli Institute of Cosmological Physics at the University of Chicago, 
the Center for Cosmology and Astro-Particle Physics at the Ohio State University,
the Mitchell Institute for Fundamental Physics and Astronomy at Texas A\&M University, Financiadora de Estudos e Projetos, 
Funda{\c c}{\~a}o Carlos Chagas Filho de Amparo {\`a} Pesquisa do Estado do Rio de Janeiro, Conselho Nacional de Desenvolvimento Cient{\'i}fico e Tecnol{\'o}gico and 
the Minist{\'e}rio da Ci{\^e}ncia, Tecnologia e Inova{\c c}{\~a}o, the Deutsche Forschungsgemeinschaft and the Collaborating Institutions in the Dark Energy Survey. 

The Collaborating Institutions are Argonne National Laboratory, the University of California at Santa Cruz, the University of Cambridge, Centro de Investigaciones Energ{\'e}ticas, 
Medioambientales y Tecnol{\'o}gicas-Madrid, the University of Chicago, University College London, the DES-Brazil Consortium, the University of Edinburgh, 
the Eidgen{\"o}ssische Technische Hochschule (ETH) Z{\"u}rich, 
Fermi National Accelerator Laboratory, the University of Illinois at Urbana-Champaign, the Institut de Ci{\`e}ncies de l'Espai (IEEC/CSIC), 
the Institut de F{\'i}sica d'Altes Energies, Lawrence Berkeley National Laboratory, the Ludwig-Maximilians Universit{\"a}t M{\"u}nchen and the associated Excellence Cluster Universe, 
the University of Michigan, the National Optical Astronomy Observatory, the University of Nottingham, The Ohio State University, the University of Pennsylvania, the University of Portsmouth, 
SLAC National Accelerator Laboratory, Stanford University, the University of Sussex, and Texas A\&M University.

The DES data management system is supported by the National Science Foundation under Grant Number AST-1138766.
The DES participants from Spanish institutions are partially supported by MINECO under grants AYA2012-39559, ESP2013-48274, FPA2013-47986, and Centro de Excelencia Severo Ochoa SEV-2012-0234.
Research leading to these results has received funding from the European Research Council under the European Union’s Seventh Framework Programme (FP7/2007-2013) including ERC grant agreements 
 240672, 291329, and 306478.

This research uses services or data provided by the NOAO Science Archive. 
NOAO is operated by the Association of Universities for 
Research in Astronomy (AURA), Inc. under a cooperative agreement 
with the National Science Foundation.

\bibliography{GW150914}

\end{document}